\documentclass{article} 
\usepackage{epsf} 
\usepackage{amsmath}
\usepackage{graphicx}
\usepackage{hyperref}

\newcommand{\MS}{\overline{\mathrm{MS}}}
\newcommand{\RI}{\mathrm{RI}}
\newcommand{\G}{\, \mathrm{GeV}}

\begin{document}
\thispagestyle{empty} \parskip=12pt \raggedbottom

\vspace*{1cm}
\begin{center}
  {\LARGE 2+1 Flavor QCD simulated in the $\epsilon$-regime in 
    different topological sectors}

  
  \vspace{1cm} 
  P.~Hasenfratz${}^a$, D.~ Hierl${}^b$, V.~Maillart${}^a$, 
  F.~Niedermayer${}^a$,
  A.~Sch\"afer${}^b$, C.~Weiermann${}^a$ and M.~Weingart${}^a$ \\ 

  ${}^a$Institute for Theoretical Physics \\
  University of Bern \\
  Sidlerstrasse 5, CH-3012 Bern, Switzerland \\
  \vspace{0.5cm}
  ${}^b$Institute for Theoretical Physics, University of Regensburg, D-93040
  Regensburg, Germany \\
  
  \vspace{0.5cm}
  
  \nopagebreak[4]
  
  \begin{abstract}
    We generated configurations with the parametrized fixed-point Dirac 
    operator $D_\mathrm{FP}$ on a $(1.6\,\mathrm{fm})^4$ box at a lattice 
    spacing $a=0.13\,\mathrm{fm}$.
    We compare the distributions of the three lowest $k=1,2,3$ eigenvalues 
    in the $\nu= 0,1,2$ topological sectors with that of the Random Matrix 
    Theory predictions. The ratios of expectation values of the lowest 
    eigenvalues and the cumulative eigenvalue distributions are studied 
    for all combinations of $k$ and $\nu$. 
    After including the finite size correction from one-loop chiral
    perturbation theory we obtained for the chiral condensate 
    in the $\MS$ scheme $[\Sigma(2\G)]^{1/3} = 0.239(11)\G$, 
    where the error is statistical only.
  \end{abstract}

\end{center}

\eject

Spontaneous chiral symmetry breaking and the related existence of light
Goldstone bosons is a basic feature of QCD. Chiral Perturbation
Theory (ChPT) provides a systematic description of this physics in terms of a
set of low energy constants which encode the related non-perturbative features
of QCD. The method of low energy effective Lagrangians simplifies the
calculations significantly \cite{SW,GL1,GL2,GL3} and over the years ChPT
became a refined powerful technique. Gasser and
Leutwyler recognized very early, decades before the numerical calculations
could attack such problems, that these constants can also be fixed using
physical quantities which, presumably, will never be measured in real
experiments. They can be studied, however, in lattice QCD.

The $\epsilon$- regime \cite{GL4,GL5,HaLe,FCH} 
describes physics close to the chiral limit in a box whose size is larger
than the QCD scale. On the other hand, the size of the box relative to the
Goldstone boson correlation length must be small. Under these conditions the
Goldstone bosons, as opposed to other excitations, feel the effect of
boundaries strongly.  ChPT provides a powerful and systematic way to calculate
the finite size corrections. A nice additional feature of the
$\epsilon$-regime is that Random Matrix Theory (RMT) \cite{rmt} 
makes precise predictions for microscopic observables. RMT relates in
particular the distribution of low-lying eigenvalues of the Dirac operator
in different topological sectors to the chiral condensate.

The pioneering numerical works \cite{EHKN,Biet1,Biet2,GLWW,Giuet,Edwards,DeG}
in the $\epsilon$-regime in {\it quenched} QCD suggested that this 
regime can be an excellent tool to study low-energy physics in QCD. The
special problems of the $\epsilon$-regime called for new numerical
procedures \cite{Giuet,Edwards,DeG,GHLW} which were first tested also in this
approximation. Combining these numerical developments with the
renormalization properties of the spectral density the work \cite{GiNe}
present a state of the art analysis for the quark condensate and the first
Leutwyler-Smilga sum rule in quenched QCD.  
The quenched approximation, however, tends to be singular in the chiral limit 
and it is expected that quenching is even more problematic here than in other
cases \cite{Dgaa}.

Unfortunately, full QCD simulations are expensive. 
In addition, in the $\epsilon-$regime the Dirac operator should have 
excellent chiral properties. 
The standard choice is the overlap Dirac operator
\cite{Neu} with a hybrid Monte Carlo algorithm.

The results in \cite{DeGSCH}
reflect already the basic physics features of the $\epsilon$-regime. 
The distribution of the low-lying eigenmodes could be fitted quite well 
to the RMT predictions and a reasonable value for the chiral condensate 
was obtained.  
These results are promising given the fact that the box was small
($1.3\,\mathrm{fm}$), the quark mass was rather large 
($m_q \ge 40\,\mathrm{MeV}$) and the lattice was coarse 
($a=0.16\,\mathrm{fm}$). 
The results in a larger box of $1.5\,\mathrm{fm}$ and 
at smaller quark mass $m_q \approx 20\,\mathrm{MeV}$ 
obtained in \cite{DeGLiuSch} were consistent with those of \cite{DeGSCH}.

The first serious simulation results in the $\epsilon$-regime
have been presented by the JLQCD group recently \cite{JLQCD}. 
In that work $N_f=2$ QCD was simulated with overlap
fermions (created with the Wilson kernel) on a lattice of 
size $1.8^3 \times 3.6\,\mathrm{fm}^4$ with a resolution $a=0.11\,\mathrm{fm}$ 
obtained from $r_0$. Different bare quark
masses ($am_q$) were considered in the range $0.110,\dots,0.020,0.002$. 
The smallest quark mass (corresponding to $\approx 3\,\mathrm{MeV}$) was
certainly small enough to reach the $\epsilon$-regime.
Hybrid Monte Carlo with overlap fermions has problems when the topological
charge changes.  
To avoid this an action was used which prevented topology change 
and the whole run stayed in the $Q=0$ sector. 
The authors observed an overall good agreement with RMT. The
fermion condensate, which is the only parameter to be fitted, was found to be 
$\Sigma(2\,\mathrm{GeV})^{1/3} = 0.251(7)(11) \mathrm{GeV}$ in 
the $\overline{\mathrm{MS}}$ scheme.

Our work is based on the parametrized fixed-point (FP) action which has been
tested in detail in quenched QCD \cite{tests}. The exact FP Dirac operator
satisfies the Ginsparg-Wilson relation 
\begin{equation} \label{DGW}
  D_\mathrm{GW}^\dagger + D_\mathrm{GW} = 
  D_\mathrm{GW}^\dagger 2R D_\mathrm{GW} \,,
\end{equation}
where $R$ is a local operator and is trivial in Dirac space.   
The parametrized FP action has many gauge paths and involves a special
smearing with projection to the gauge group $\mathrm{SU}(3)$. 
As a consequence, hybrid
Monte Carlo algorithms can not be used. In the work \cite{APF} we advised a
partially global update with three nested accept/reject steps which can
reach small quark masses even on coarse lattices. Several steps of this
algorithm were developed in \cite{AHFK} using  some earlier
suggestions \cite{Gr,MC,AHAA,KW}.  All the three steps are preconditioned.
Pieces of the quark determinant are switched on gradually in the order of
their computational expenses. 
The largest and most fluctuating part of the determinant is
coming from the ultraviolet modes. 
We reduce these fluctuations by calculating the trace of
$D_\mathrm{FP}^n, n=1,...,4$ \cite{MH2001,PdeF,AHFK}. 
The $\sim 100$ lowest-lying modes are calculated and subtracted. 
The determinant of the reduced and subtracted $D_\mathrm{FP}$ is 
calculated stochastically \cite{KK,KKMP,Joo,MSCH}. 
For the determinant breakup we generalized the mass shifting method of
\cite{MH2001}. For the strange quark we performed a root operation.
Since the strange quark mass is not very small, we used a polynomial expansion
to approximate this root operator. 
Recently we replaced the polynomial expansion by a rational approximation 
\cite{Manuel} which brought a $\approx 30\%$ performance gain in this part.

Due to the partial global updating procedure the algorithm, beyond a certain
volume, scales with $V^2$ which constrains the size of lattices which can be
considered. 
Due to the large number of subtracted low-lying modes, small quark mass is 
not a barrier.

We generated $\sim 4000$ configurations in the Markov chain
on an $12^4$ lattice using the partially global algorithm discussed above. 
The distance between every second configurations in this chain is similar 
to that of two gauge configurations separated by a typical Metropolis 
gauge update sweep. 
We considered every tenth configurations from
the Markov chain and performed the measurements on the remaining 
$\sim 400$ configurations. When calculating statistical errors we formed 20
bins and used the jackknife method.

We fixed the lattice spacing from the Sommer parameter \cite{Sommer} 
$r_0=0.49\,\mathrm{fm}$ and found $a=0.129(5)\,\mathrm{fm}$. 
Our box has the size $(1.6\,\mathrm{fm})^4$.

\begin{figure}[ht]
  \begin{center}
    \includegraphics[width=100mm]{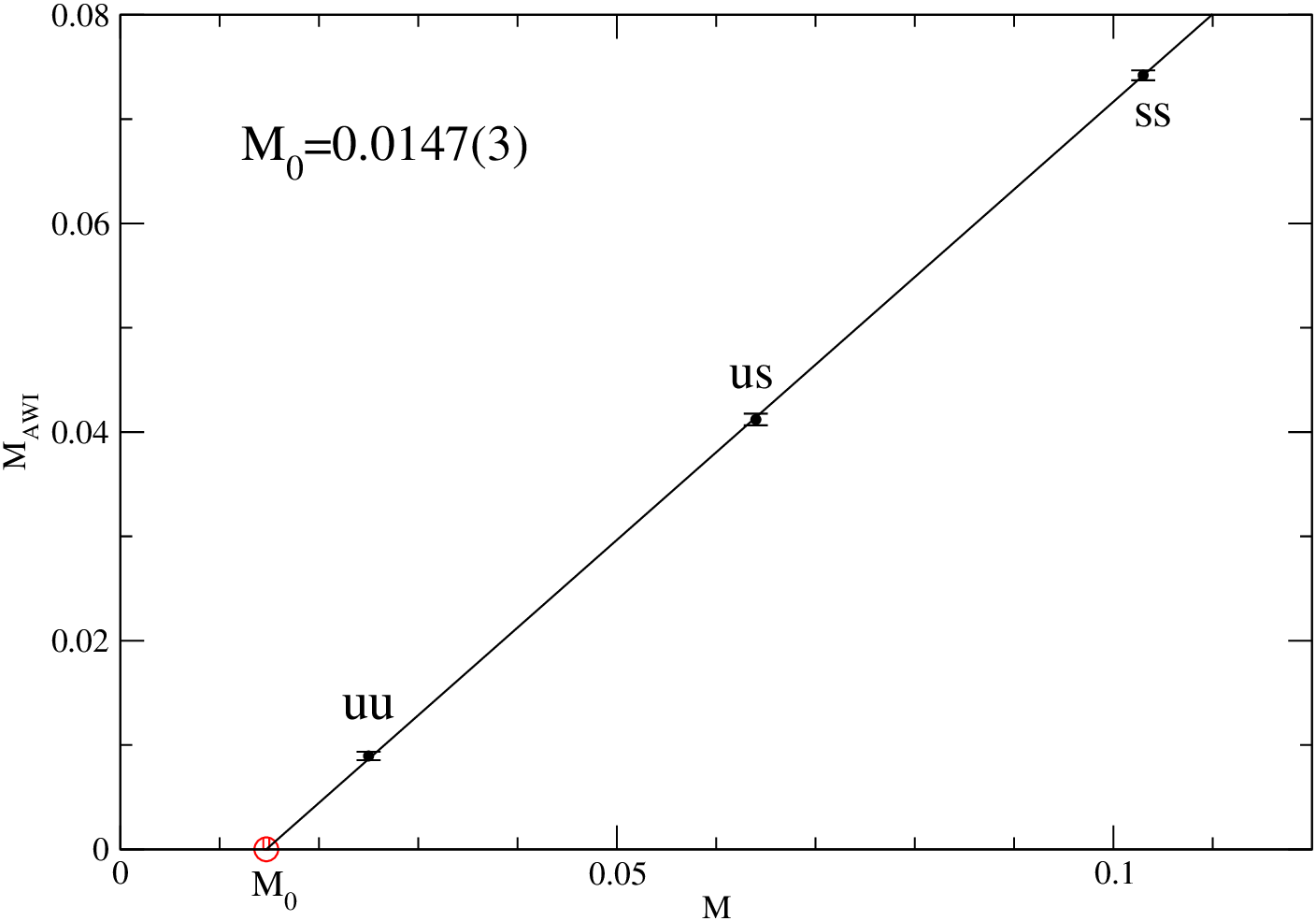}
    \caption{{} The AWI mass for the three combinations of quark masses.
      The linear extrapolation to $M_\mathrm{AWI}=0$ gives an additive 
      mass renormalization $M_0=0.0147(3)$.
    } \label{fig:awi} 
  \end{center}
\end{figure}

The Dirac operator $D_\mathrm{FP}$ has no exact chiral symmetry due to
parametrization errors. As a consequence, the quark masses have an 
additive mass renormalization. 
The degenerate $u,\, d$ and the $s$ quark masses in the 
code are $M_\mathrm{ud} = 0.025$ and  $M_\mathrm{s} =0.103$. 
The additive renormalization was measured following the steps 
described in \cite{Dieter} 
and has the value $M_0=0.0147(3)$, as shown in Fig.~\ref{fig:awi}.
Subtracting the additive renormalization we get the bare masses
$m_\mathrm{ud}=0.0103(3)$ and $m_\mathrm{s}=0.0883(3)$ which corresponds to 
$16\,\mathrm{MeV}$ and $137\,\mathrm{MeV}$, respectively. 

\begin{figure}[ht]
  \begin{center}
    \includegraphics[width=100mm]{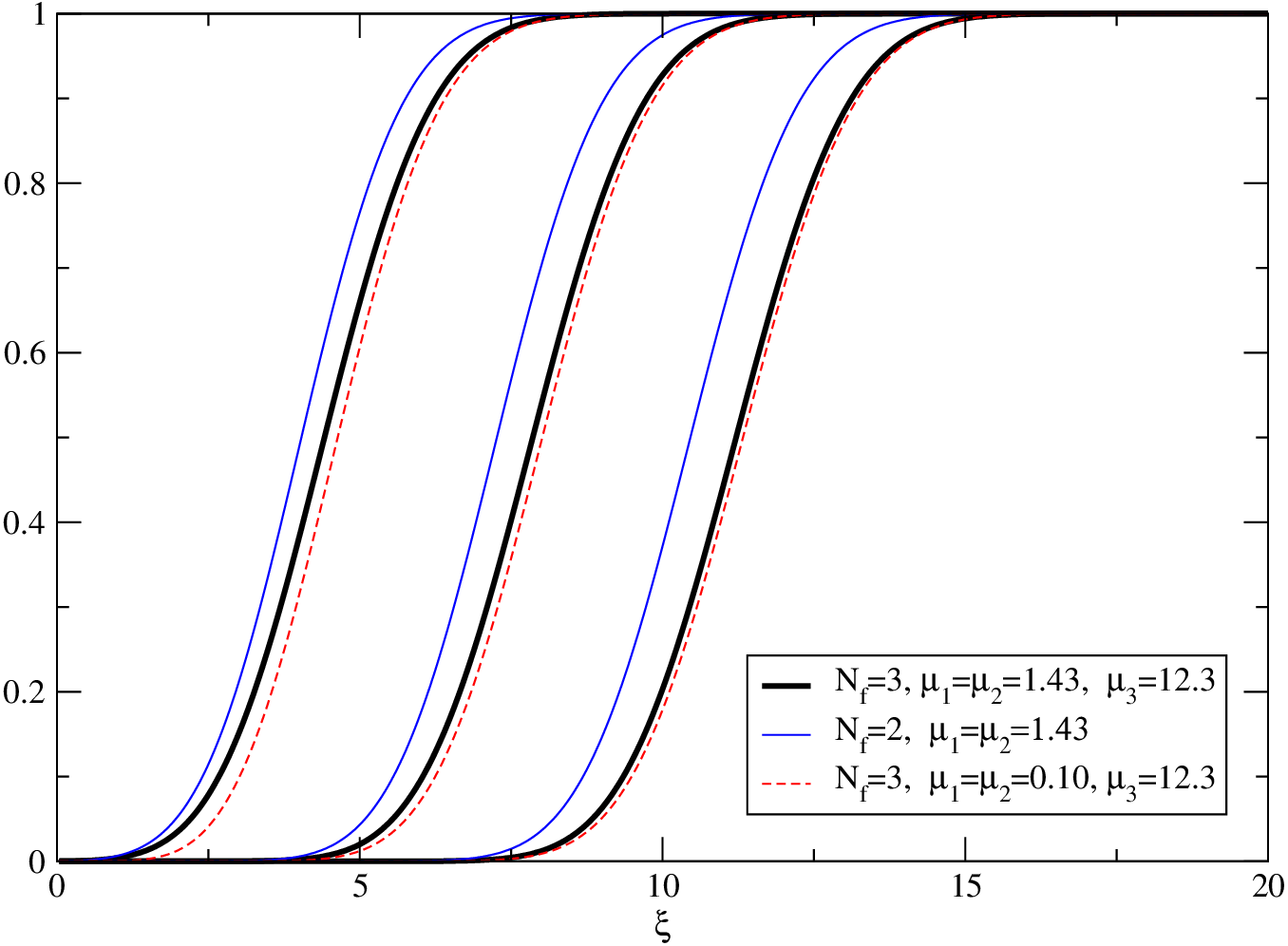}
    \caption{{} Random Matrix Theory prediction for the cumulative 
      distribution of $\xi = \xi_{\nu k} = \alpha_{\nu k} \Sigma V$, 
      where $i \alpha_{\nu k}$ is the k-th eigenvalue of the continuum 
      Dirac operator in a gauge background with topological charge 
      $\nu$, and $\mu_i= m_i \Sigma V$.
      Here the $\nu=0$ results are shown, but the picture is similar
      for $\nu=1,2$.} \label{fig:rmt}  
  \end{center}
\end{figure}

As we mentioned above, a simulation with lighter quark masses would not be
more expensive. 
In our case, however, a smaller $m_\mathrm{ud}$ were not useful 
either. RMT predicts the probability distribution $p_{\nu k}(\xi_{\nu k})$ 
for the $k$-th low-lying eigenvalue ($ k=1,2,\dots$) of the Dirac operator 
in the topological sector $\nu$.  
Denoting the corresponding eigenvalues of the (continuum) Dirac operator
by $i \alpha_{\nu k}$, the variable $\xi_{\nu k}$ is related to 
the bare chiral condensate $\Sigma$ as 
$\xi_{\nu k}= \alpha_{\nu k}\Sigma V$. 
Here V is the volume of the box. 
Fig.~\ref{fig:rmt} shows the prediction of RMT for the cumulative 
distributions $\int^{\xi_{\nu k}}_0 d\xi p_{\nu k}(\xi)\,$ for 
$\nu=0$ and $k=1,2,3$. The distributions depend on $\mu_i=m_i \Sigma V$ 
where $m_i$ are the quark masses. 
Decreasing the $m_\mathrm{ud}$ mass by more than a factor 10 
(at fixed $m_\mathrm{s}$)
the cumulative distributions practically remain unchanged. On the other hand
sending the strange quark mass to infinity (at fixed $m_\mathrm{ud}$) 
we land on a $N_f=2$ flavor theory with a visibly different distribution. 
The strange quark has a (modest) effect on our observables.

\begin{figure}[ht]
  \begin{center}
    \includegraphics[width=100mm]{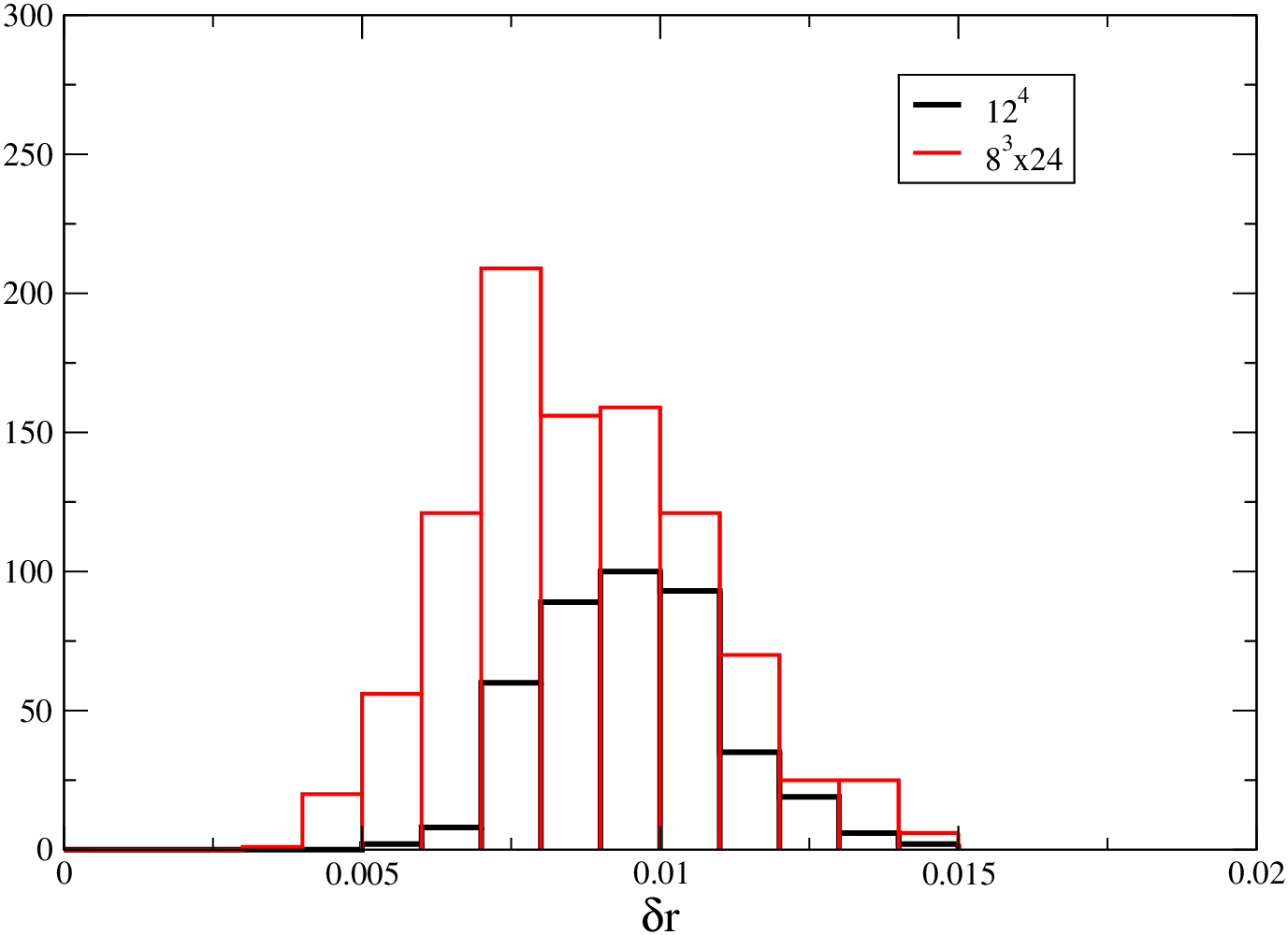}
    \caption{{} The histogram of the deviation from the circle,
      $\delta r = 1 - |\lambda-1|$, for the eigenvalues of $D_\mathrm{FP}$
      on $8^3\times 24$ and $12^4$ lattices for the complex
      eigenvalues with $|\mathrm{Im}\lambda| < 0.1$, using the present
      mass parameters.} \label{fig:dr}  
  \end{center}
\end{figure}

While the eigenvalues of the continuum Dirac operator lie on the imaginary
axis the spectrum of lattice Dirac operators is more complicated.
As it is well known, the spectrum of the Dirac operator satisfying
the Ginsparg-Wilson relation with $2R=1$ lies on the circle
$|\lambda-1|=1$.  
In our case where $2R$ is different from 1, it is convenient
to introduce the rescaled operator
$\hat{D}_\mathrm{GW} = \sqrt{2R} D_\mathrm{GW} \sqrt{2R}$,
for which the $2R$ factor is eliminated in the GW relation,
and the spectrum lies on the circle.
(In fact, our operator $2R$ for the low-lying
modes is effectively a constant close to 1 within a few percent.)
To relate the eigenvalues on the GW circle to those appearing
in the RMT (or in general, in continuum expressions) it is natural
to use the stereographic projection 
\begin{equation}
  i\alpha = \frac{\lambda}{1-\lambda/2} \,. 
\end{equation}
As Fig.~\ref{fig:dr} shows, the low-lying eigenvalues of $D_\mathrm{FP}(m)$ 
(which we obtain
during our simulation) are close but not exactly on the GW circle,
hence we project them first horizontally onto the GW circle,
and make the stereographic projection in the next step.
This is justified by the observation that making a systematic
GW improvement of $D_\mathrm{FP}$ towards $D_\mathrm{GW}$ 
the imaginary part of the eigenvalues stays practically constant --
they move horizontally to the GW circle. 
The values of $\alpha$ obtained by this procedure are then 
compared with the RMT predictions.

\begin{figure}[ht]
  \begin{center}
    \includegraphics[width=100mm]{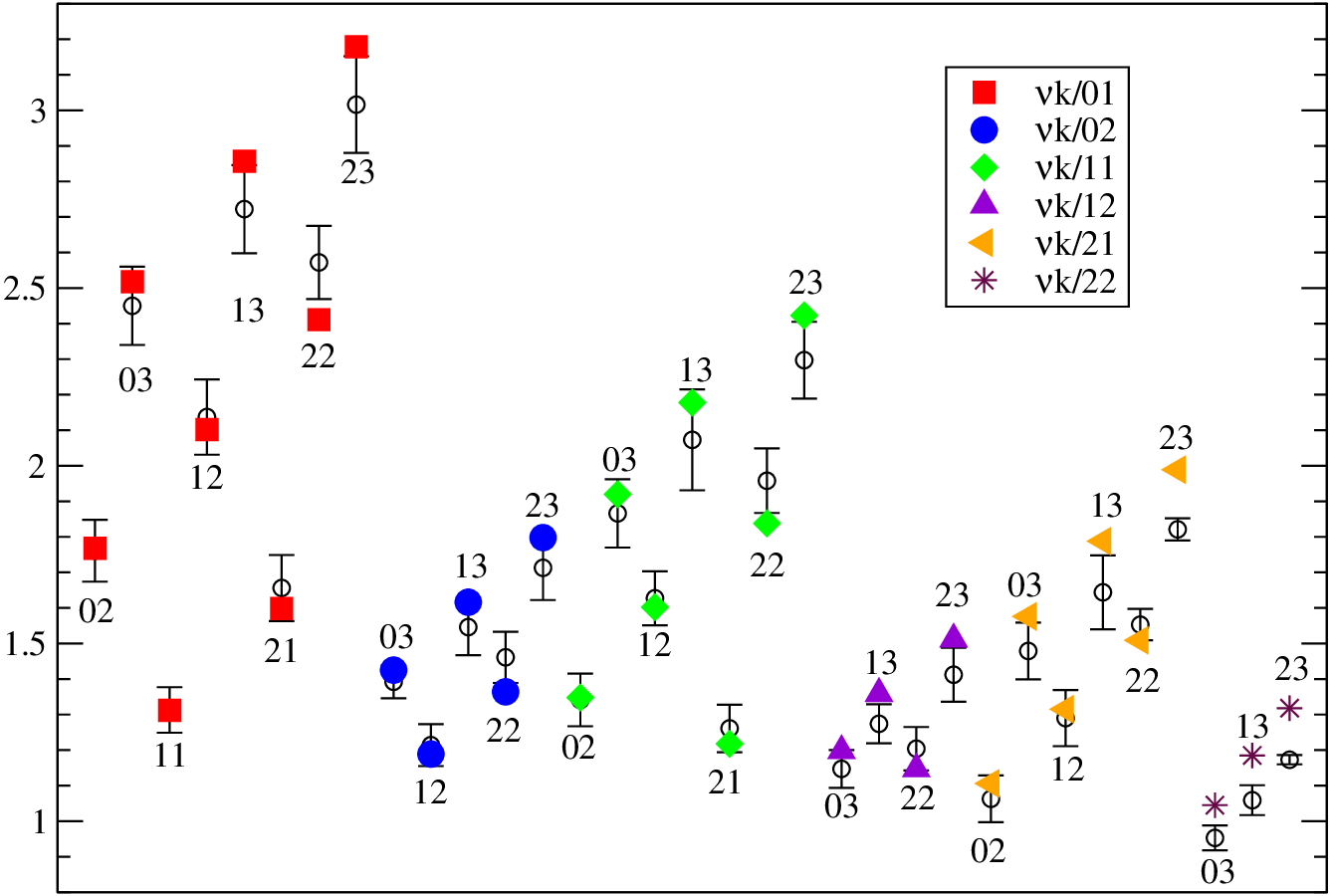}
    \caption{{} Ratios of expectation values 
      $\langle \alpha_{\nu k} \rangle$ obtained from simulations
      compared to the results of RMT, where $\nu k$ is indexing 
      the $k$-th lowest
      eigenvalue in the topological sector $\nu$. The different symbols 
      refer to the denominator while $\nu k$ of the numerator is 
      indicated at the data points. For example, the highest ratio with value
      $\approx 3$ in the figure refers to 
      $\langle \xi_{23} \rangle / \langle \xi_{01} \rangle $ }.
    \label{fig:ratios}  
  \end{center}
\end{figure}

Having the probability distributions from RMT we can calculate the ratios 
$\langle \xi_{\nu k} \rangle / \langle \xi_{\nu' k'} \rangle $, where the
factor $\Sigma V$ cancels. These predictions are compared with
the measured ratios in Fig.~\ref{fig:ratios}, where all the ratios 
in the topological sectors $\nu=1,2,3$ for the first three lowest 
eigenvalues are shown.

\begin{figure}[ht]
  \begin{center}
    \includegraphics[width=100mm]{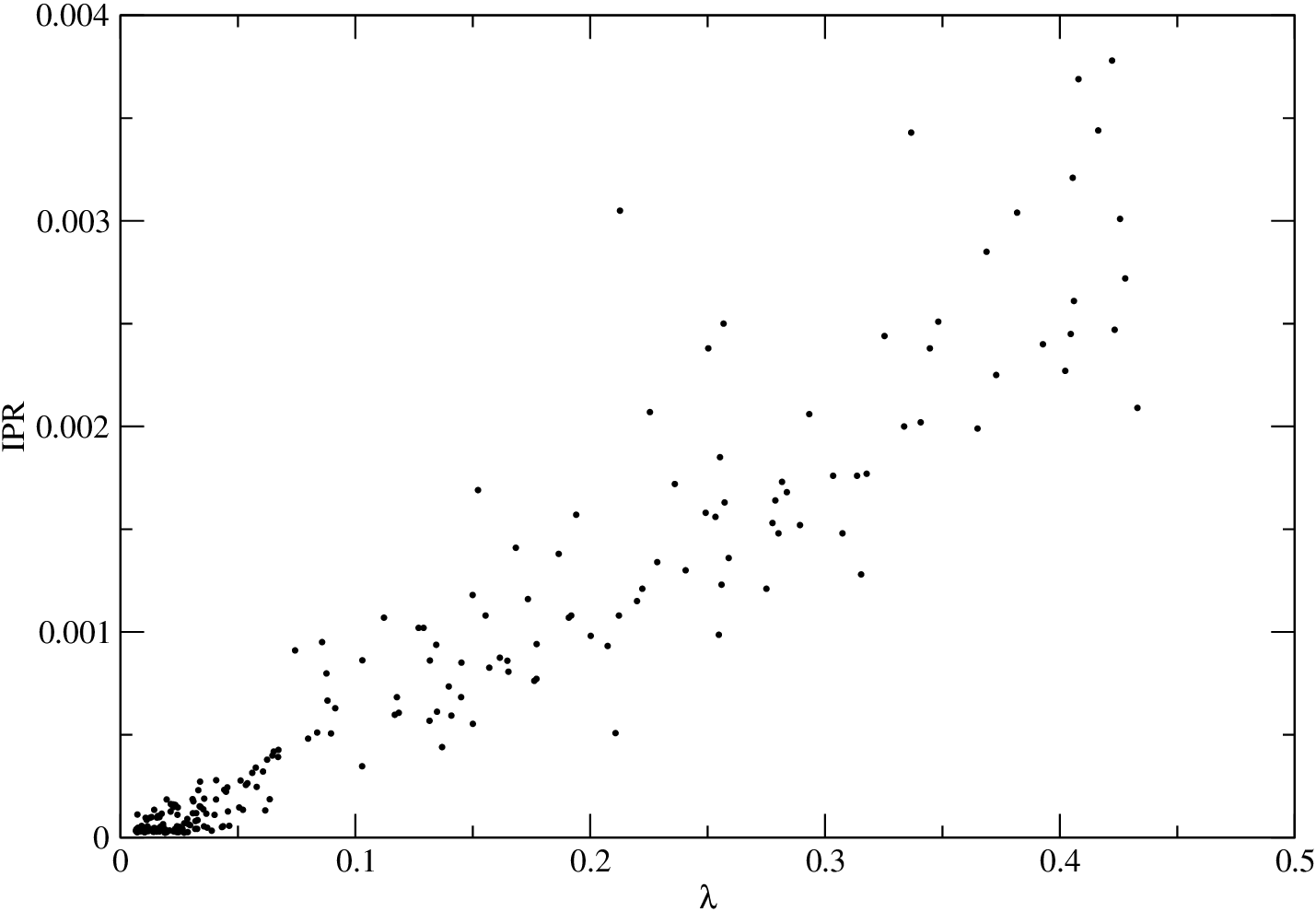}
    \caption{{} The inverse participation ratio 
      $\sum_x \sum_{i=1}^{12} |\psi_i^{(\lambda)}(x)|^4$
      for {\em real} eigenvalues $\lambda(D_\mathrm{FP})$ vs. the eigenvalue. 
      Eigenvectors with larger real part have smaller 
      extension in agreement with the expectations.} \label{fig:IPR} 
  \end{center}
\end{figure}

\begin{figure}[ht]
  \begin{center}
    \includegraphics[width=100mm]{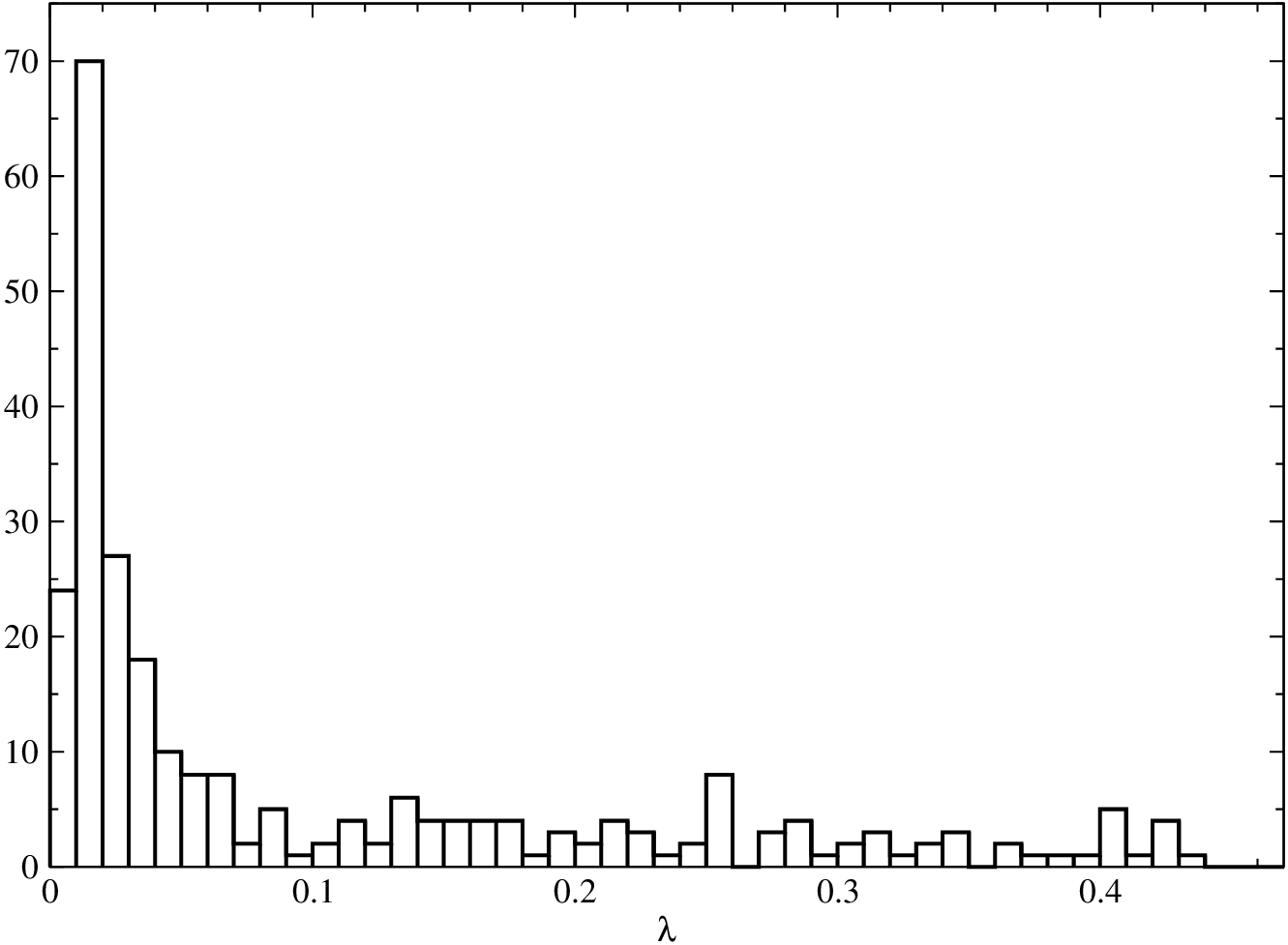}
    \caption{{} Histogram of $\lambda(D_\mathrm{FP})$
      for the real eigenvalues. In determining the topological charge
      $\nu$ we considered only the real eigenvalues with
      $\lambda < 0.03$.} \label{fig:hist_Re_ev} 
  \end{center}
\end{figure}

At this point we have to discuss the way we identify the topological
sectors. According to the index theorem \cite{HLN} one can identify the
topological charge from the zero modes of a Ginsparg-Wilson Dirac
operator. However, our $D_\mathrm{FP}$ has no exact chiral symmetry 
and has modes on the real axis. 
Occasionally, the real eigenvalue $\lambda$ might be even far
from zero in which case the topological interpretation is uncertain. 
We note here that using an exact Ginsparg-Wilson operator (with kernel
$D_\mathrm{FP}$) for measuring the
topological charge is not a good solution to this problem.
The overlap just projects most of the real eigenvalues to the 
point $\lambda=0$. 
Following the intuitive picture that topology is related 
to  extended objects we investigated the correlation between 
the inverse participation ratio (IPR) of the eigenvector having a real 
eigenvalue $\lambda$ and the value of
$\lambda$. Here IPR is given by 
$\sum_x \sum_{i=1}^{12} |\psi_i^{(\lambda)}(x)|^4$ for a 
normalized eigenvector $\psi^{(\lambda)}$. 
The IPR is inversely proportional to the size of the effective support of
the eigenfunction. 
As Fig.~\ref{fig:IPR} shows, there is a strong correlation indeed. 
As $\lambda$ is moving away from zero, the IPR increases
indicating that the wave function becomes more and more localized. 
Fig.~\ref{fig:hist_Re_ev} demonstrates that the real eigenvalues 
$\lambda(D_\mathrm{FP})$ are strongly concentrated in a small region 
close to zero. 
The histogram of $\lambda(D_\mathrm{FP})$ 
has a large, narrow peak followed by a long tail.

\begin{figure}[ht]
  \begin{center}
    \includegraphics[width=100mm]{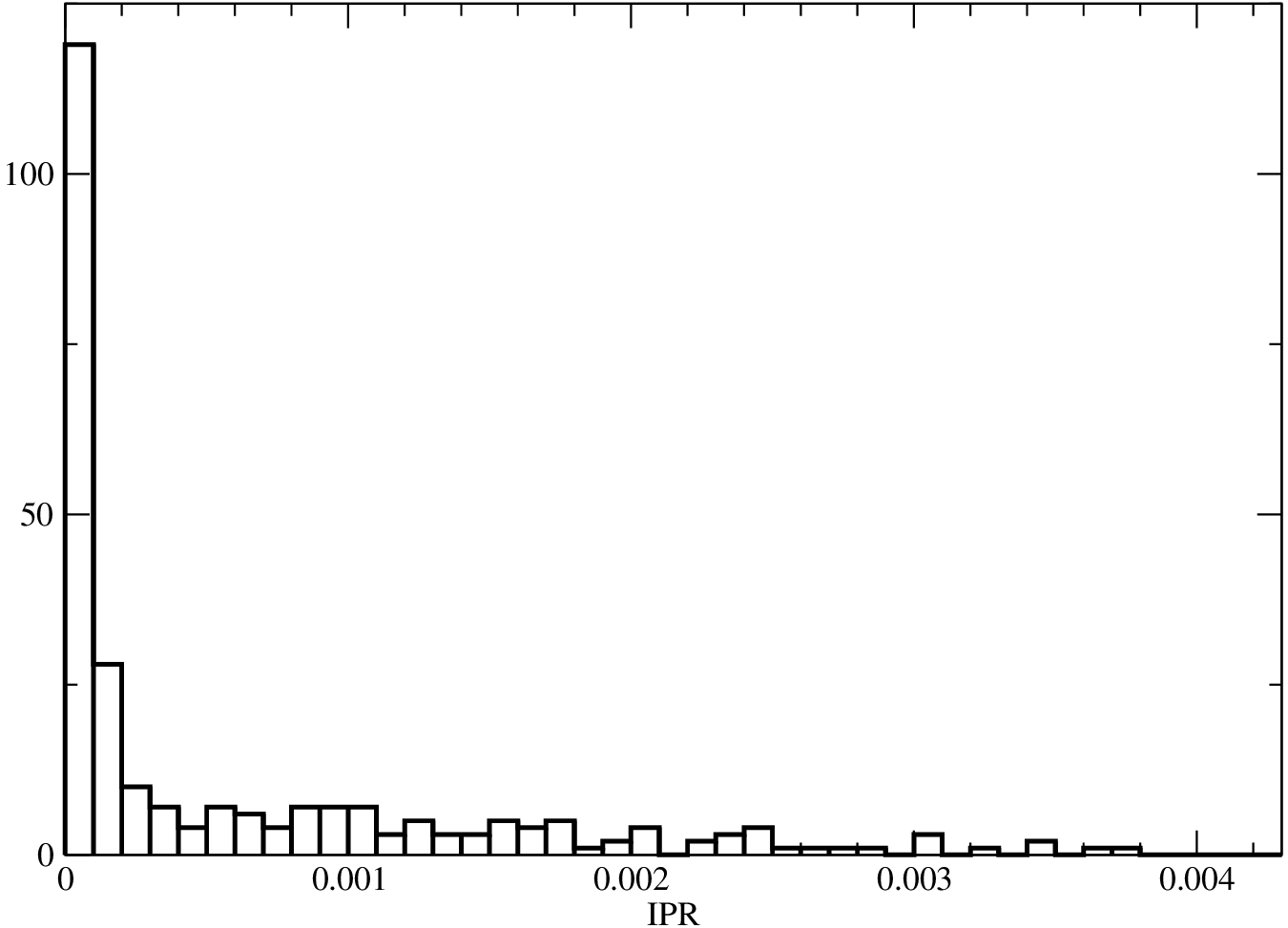}
    \caption{{} Histogram of inverse participation ratio
      for the real eigenmodes of $D_\mathrm{FP}$.} \label{fig:hist_IPR} 
  \end{center}
\end{figure}

Fig.~\ref{fig:hist_IPR} shows the histogram of IPR for the real eigenmodes of 
$D_\mathrm{FP}$. There is a strong, narrow peak which corresponds to extended
real modes and a long tail corresponding to wave functions with smaller
support.

\begin{figure}[ht]
  \begin{center}
    \includegraphics[width=100mm]{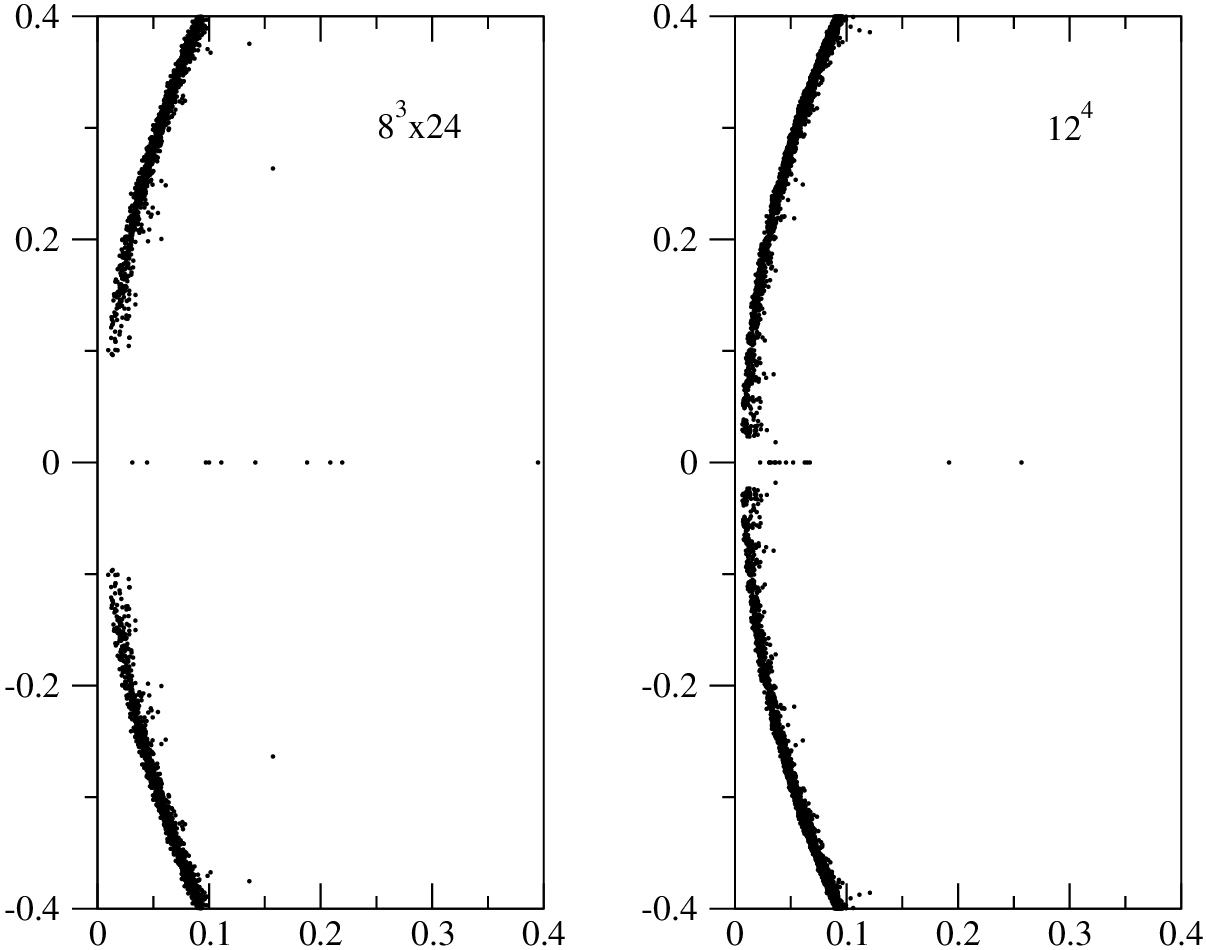}
    \caption{{} The spectrum of $D_\mathrm{FP}$ for 50 configurations
      on $8^3\times 24$ and $12^4$ lattices, at the same bare mass parameters
      and gauge coupling.} \label{fig:spectrum} 
  \end{center}
\end{figure}

For the questions we study in this paper it is not necessary to use a procedure
which leaves the relative weights of the different topological sectors intact.
(Due to the long autocorrelation time of the topological charge it
would be anyhow difficult to control these weights.) 
We introduce therefore a cut above which the real eigenvalues are
discarded. This cut must be sufficiently 
small to allow for a good identification of the topological charge, 
but large enough to allow for a
reasonable number of configurations in each sectors we want to investigate.
The cut should be smaller than the spectral gap (for the spectrum see 
Fig.~\ref{fig:spectrum}), otherwise the effect of topology will be 
strongly reduced. We introduced a cut of size 0.03 which is approximately 
equal to the minimal gap in the $\nu=1$ sector. We accepted a real
eigenvalue $\lambda$ as a sign of topology if $\lambda < 0.03$. 
This cut splits our sample of 414 configurations into 
328, 51 and 35 configurations for the $\nu=0,1,2$ sectors, respectively. 
This cut corresponds also to the end of the peak of the IPR
distribution shown in Fig.~\ref{fig:hist_IPR}.

\begin{figure}[ht]
  \begin{center}
    \includegraphics[width=100mm]{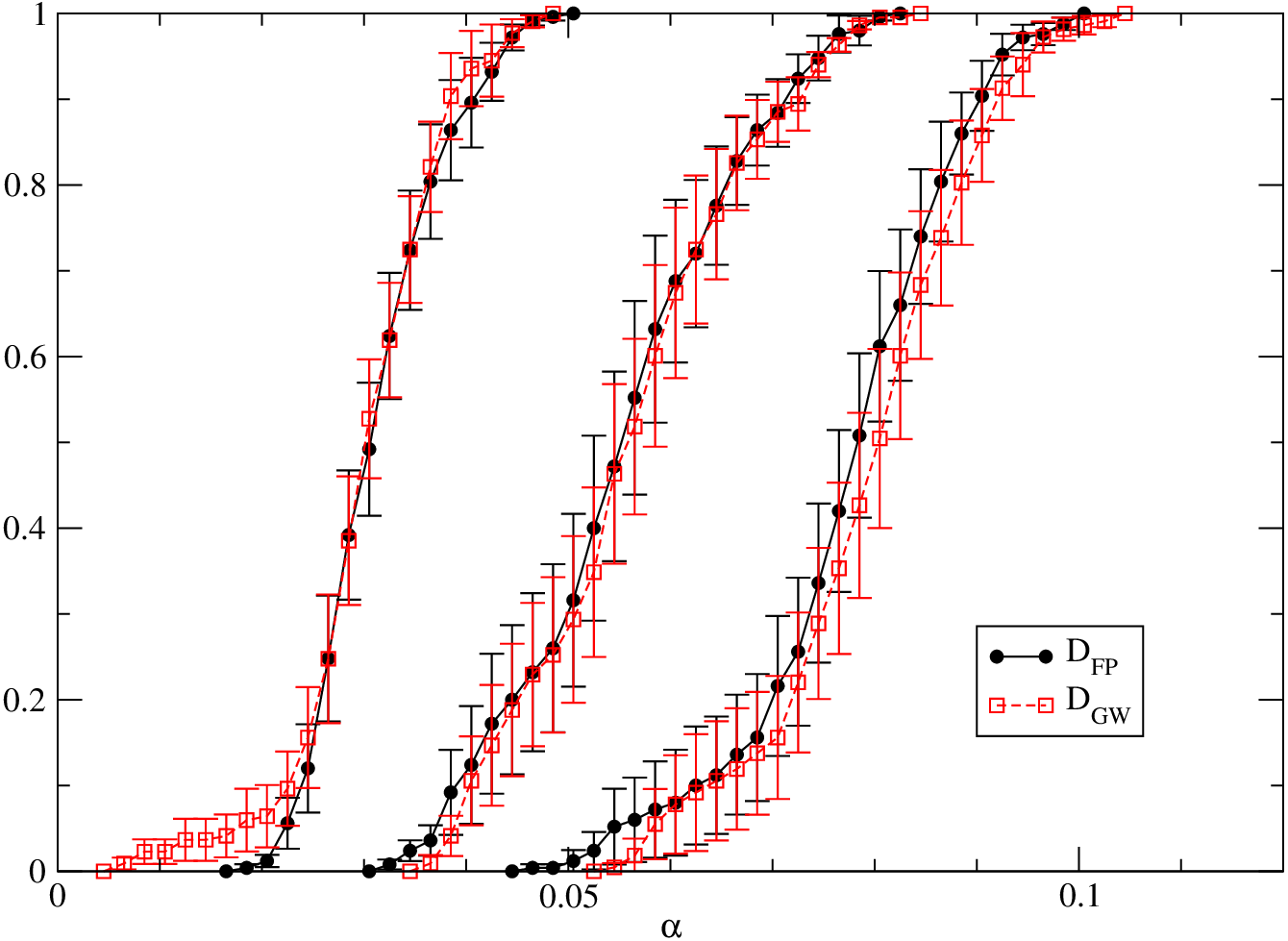}
    \caption{{} The cumulative distributions in the $\nu=0$ topological sector
      for the three lowest complex eigenvalues as measured with
      $D_\mathrm{FP}$ and with the overlap operator $D_\mathrm{GW}$ using 
      the fixed-point operator as kernel.}  \label{fig:cum_0e} 
  \end{center}
\end{figure}

Fig.~\ref{fig:cum_0e} compares the cumulative distributions of the first
three eigenvalues in the $\nu=0$ topological sector obtained using 
the overlap improved fixed-point Dirac operator $D_\mathrm{GW}$ vs. 
$D_\mathrm{FP}$ in the measurement. 
The distributions are very similar showing
that the two operators are close to each other. The $k=1$ distribution from
$D_\mathrm{GW}$ has a tail at small eigenvalues 
(i.e. has a smaller gap than $D_\mathrm{FP}$)
demonstrating a systematic error one obtains when using different Dirac 
operators for the generation of the configurations and for the measurements: 
the small eigenvalues are less effectively suppressed. 

\begin{figure}[ht]
  \begin{center}
    \includegraphics[width=100mm]{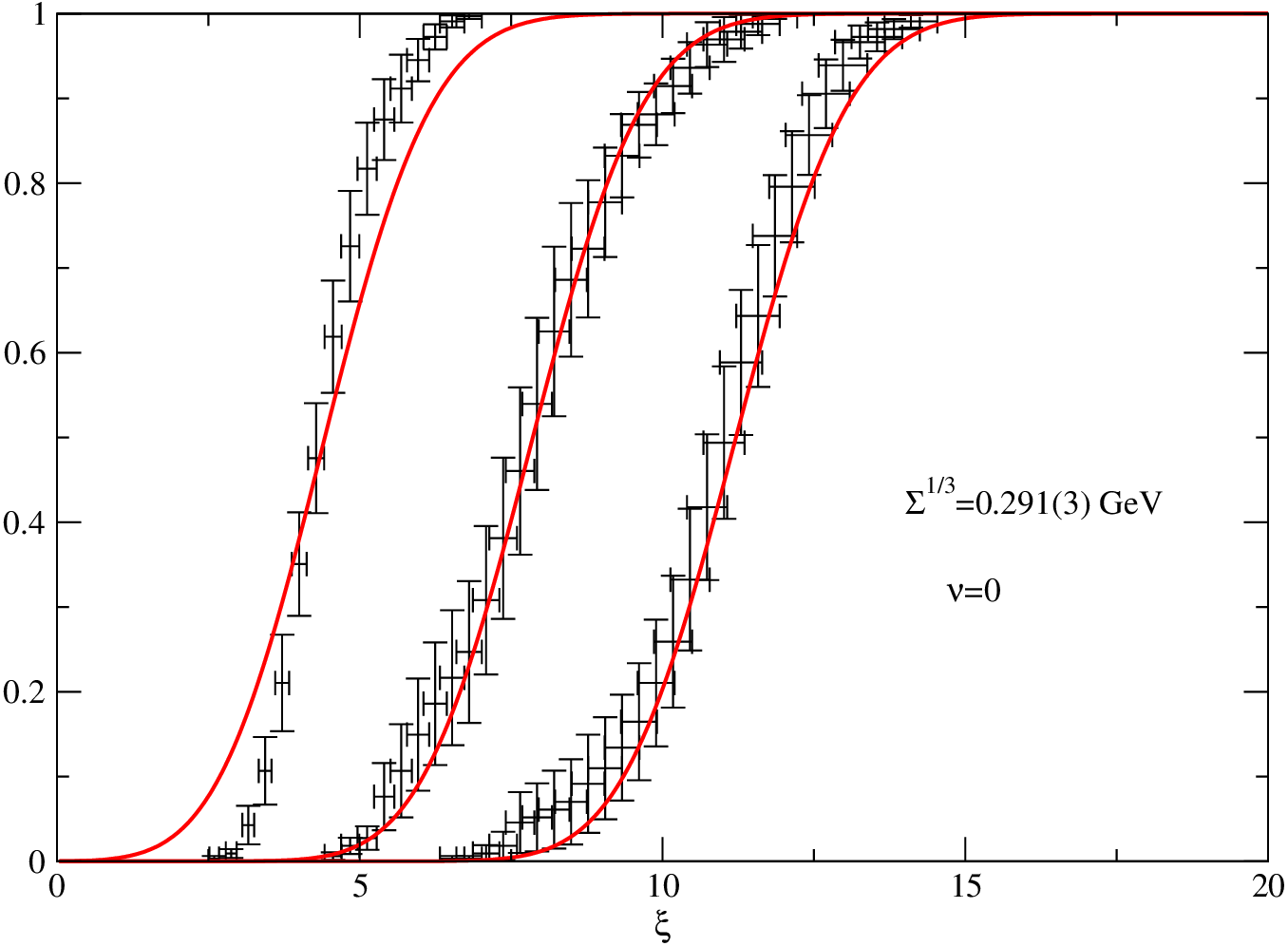}
    \caption{{} The cumulative distribution of 
      $\xi_{\nu k} = \alpha_{\nu k} \Sigma V$ for $\nu=0$, $k=1,2,3$.
      Here $\alpha$ is obtained from $\lambda(D_\mathrm{FP})$ 
      by stereographic projection described in the text, and is 
      rescaled using $\Sigma^{1/3}=0.291(3)\,\mathrm{GeV}$.
      The continuous line is the RMT prediction at
      $\mu_1=\mu_2=m_\mathrm{ud} \Sigma V = 1.43$ and
      $\mu_3=m_\mathrm{s} \Sigma V = 12.3$.
    } \label{fig:cum_0ee} 
  \end{center}
\end{figure}

\begin{figure}[ht]
  \begin{center}
    \includegraphics[width=100mm]{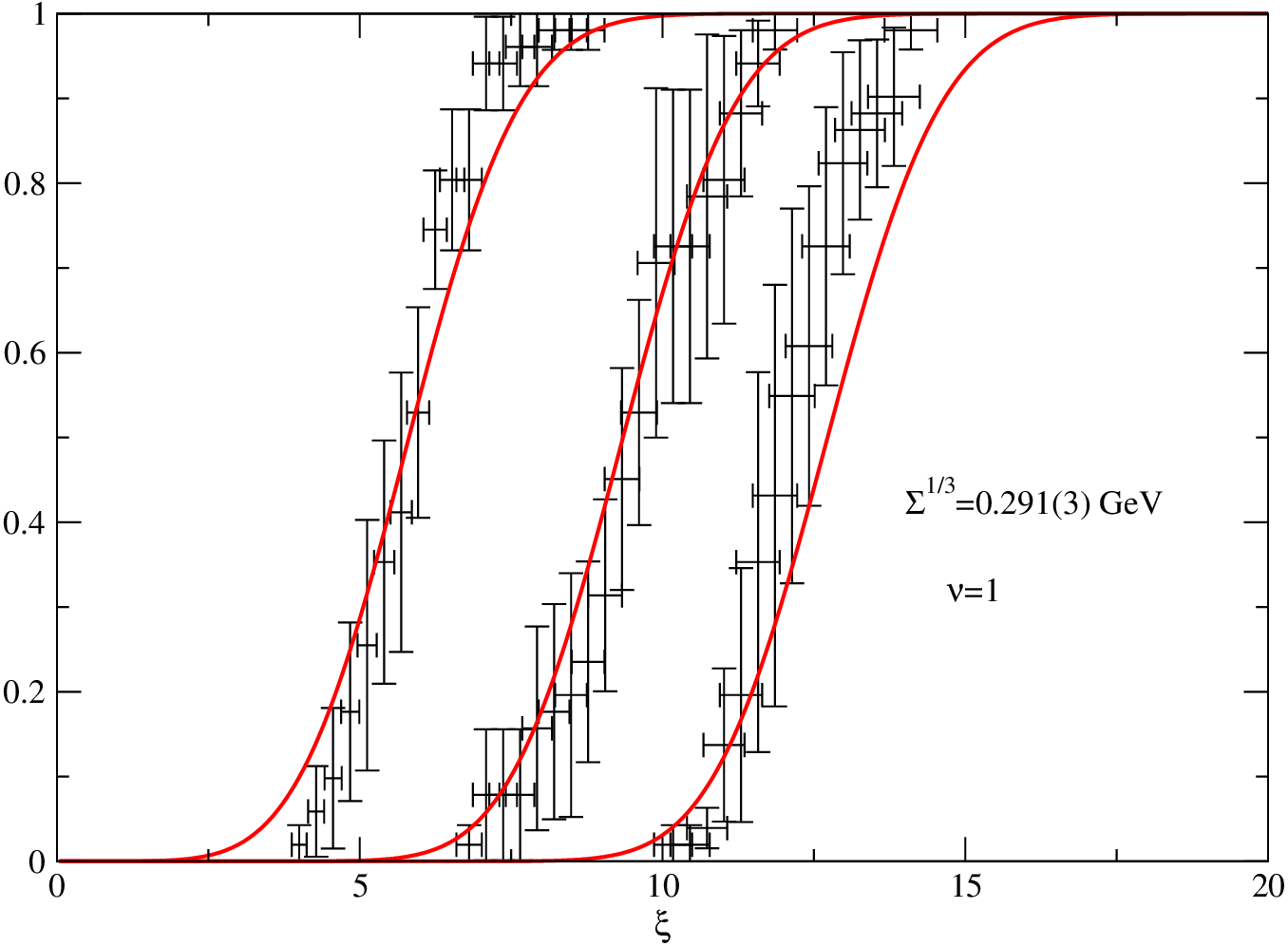}
    \caption{{} The cumulative distribution of 
      $\xi_{\nu k} = \alpha_{\nu k} \Sigma V$ for $\nu=1$, $k=1,2,3$.
    } \label{fig:cum_1ee} 
  \end{center}
\end{figure}

\begin{figure}[ht]
  \begin{center}
    \includegraphics[width=100mm]{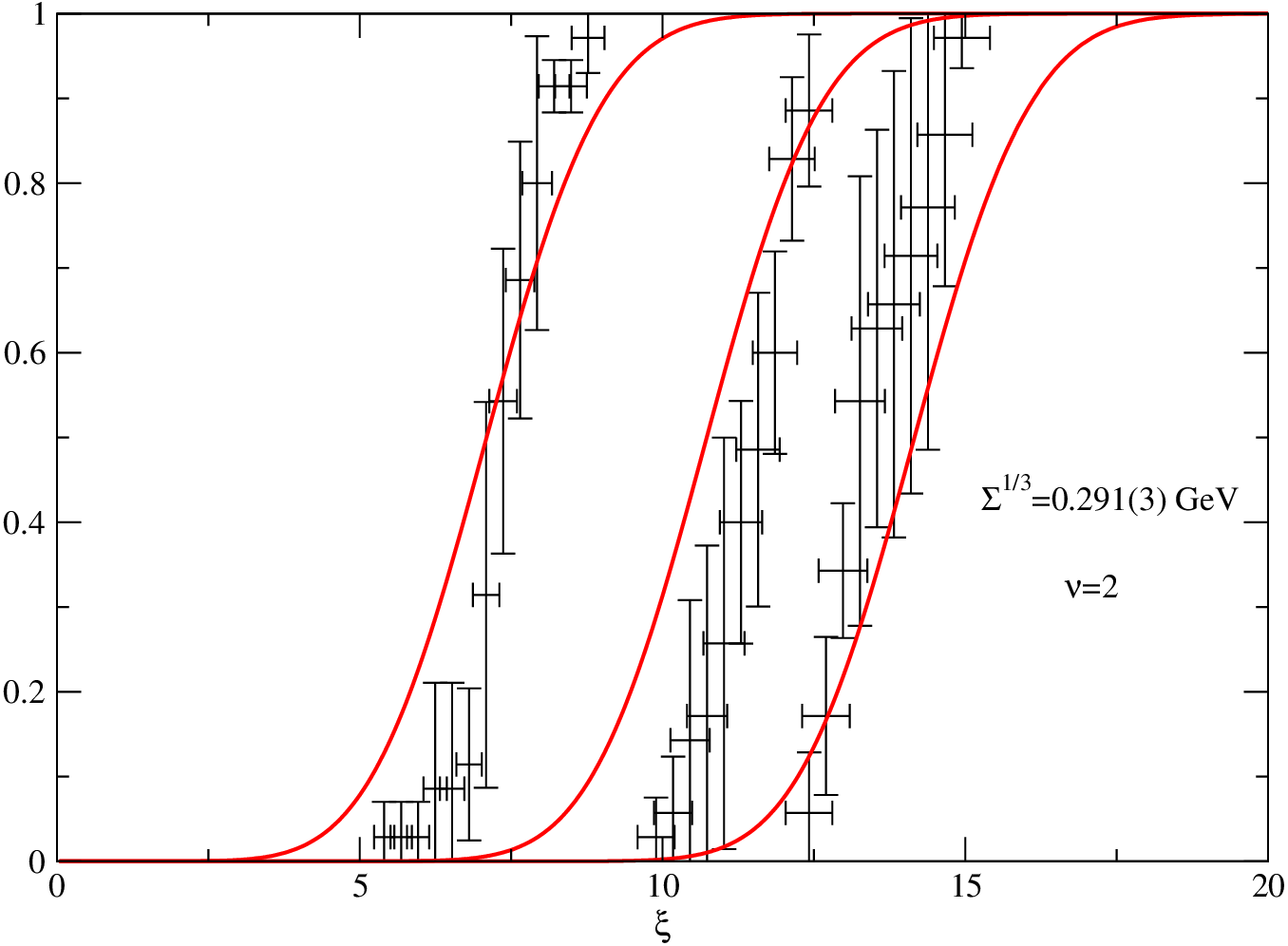}
    \caption{{} The cumulative distribution of 
      $\xi_{\nu k} = \alpha_{\nu k} \Sigma V$ for $\nu=2$, $k=1,2,3$.
    } \label{fig:cum_2ee}  
  \end{center}
\end{figure}

In Figs.~\ref{fig:cum_0ee}, \ref{fig:cum_1ee}, \ref{fig:cum_2ee} 
the cumulative distributions of the $D_\mathrm{FP}$ 
operator are compared with RMT for the $\nu=0,1,2$ topological sectors.
The only matching parameter is the bare 
condensate $\Sigma$ which enters the RMT predictions
through $\mu_i= m_i \Sigma V$ and in $\alpha \Sigma V$. We obtained the
result for this bare quantity $\Sigma^{1/3} = 0.291(3)(9)\,\mathrm{GeV}$. 
Both errors are statistical. The first error comes from the statistical 
errors of the measured distributions, the second one is due to the error 
in the lattice scale $a$. 

The result above receives a small correction due to the fact that, for
simplifying the presentation, we suppressed a technical complication. As
mentioned before the exact fixed-point operator satisfies a Ginsparg-Wilson
relation with a local operator $2R$. For this reason the quark mass
enters in a simple additive way in the form $\hat{D} + m(1-\hat{D}/2)$, where
$\hat{D}=\sqrt{2R} D_\mathrm{FP} \sqrt{2R}$.
Effectively the operator $2R$ behaves in the infrared like a 
constant close to 1. Its expectation value for the lowest
$\sim 100$ eigenvectors is $1.05$ within 1\%.
Using the spectrum of $\hat{D}$ (which is in fact 
identical to that of $2R D_\mathrm{FP}$) the matching with RMT gives 
the slightly changed result $\Sigma^{1/3} = 0.286(3)(9)\,\mathrm{GeV}$.

In the language of a ferromagnetic $\mathrm{O}(N)$ model one should 
interpret $\Sigma$ as the {\em absolute value} of the magnetization
in the finite volume $V$. 
This differs from the value $\Sigma_\infty$ defined in the infinite
volume by a finite-size correction which can be calculated 
in ChPT.  
In the presence of the magnetic field $h$ the orientation of magnetization
is controlled by the Boltzmann factor $\exp(h M \cos \theta)$ 
where $M$ is the total magnetization. This gives 
\begin{equation}
  \langle M_\parallel \rangle = M \frac{Y'_N(hM)}{Y_N(hM)} \,,
\end{equation}
where $Y_N(z)$ is related to the modified Bessel functions.
Comparing this expression with the result of ChPT
\cite{HaLe} we get for $N=4$ (corresponding to two flavors) 
$\Sigma = \rho \Sigma_\infty$, where
\begin{equation}
  \rho = \left( 1 + \frac{3}{2} \frac{\beta_1}{F^2 L^2} \right) \,.
\end{equation}
Here the shape coefficient $\beta_1$ takes the value 0.14046 for a symmetric
box \cite{HaLe}. The one-loop finite-size correction to the order parameter
(magnetization in the example above) has
been calculated for the $\mathrm{SU}(N_f) \times \mathrm{SU}(N_f)$ 
symmetry group in \cite{GL4,GL5}, for the $\mathrm{O}(N)$ group 
in \cite{HNeu1988} and up to the two-loop level in \cite{HaLe}.   

Neglecting the strange quark contribution in the finite size effects we use
the two-flavor result to 
correct the measured bare $\Sigma^{1/3} = 0.286(3)(9)\,\mathrm{GeV}$ 
by the one-loop finite size correction to obtain 
$\Sigma_\infty^{1/3} = \Sigma^{1/3}/1.119 = 0.255(3)(9)\,\mathrm{GeV}$.

In the previous version of this paper we quoted only this bare value.
Since then we have calculated the renormalization factor connecting
the bare lattice result to the $\MS$ scheme. The main points of this
calculation are summarized in the Appendix.
With the conversion factor $0.82(3)$ we obtain
\begin{equation}
  [\Sigma_{\MS}(2\G)]^{1/3} = 0.239(11)\G \,.
\end{equation}

\clearpage

{\bf Acknowledgements}
We thank Gilberto Colangelo, Stephan D\"urr, J\"urg Gasser, Anna Hasenfratz
and the members of the BGR Collaboration for valuable discussions. 
We also thank the LRZ in Munich and CSCS in Manno for support.
The analysis was done on the PC clusters of ITP in Bern.
This work was supported by the Schweizerischer Nationalfonds.

\renewcommand{\theequation}{A-\arabic{equation}}
\setcounter{equation}{0}
\section*{APPENDIX}

The bare lattice scalar density is not universal, it depends on the lattice
action and it needs renormalization. The conventional way is to express the
result in $\MS$ scheme at some given scale, e.g. $\mu_0 = 2 \G$.  

To relate the bare lattice result to $\MS$ scheme one generally uses the
Rome-Southhampton RI/MOM method \cite{Martinelli:1994ty}. In this method one
introduces an intermediate scheme  $\RI$ (or $\RI'$\footnote{These schemes
  differ in their definition of the quark field renormalization  
factor $Z_q$.}) where some Green's functions of the actual operator
(calculated in a fixed gauge) are equated to their corresponding Born terms at
a given scale $p^2 = \mu^2$.  

In the first step one relates the bare lattice operator non-perturbatively to
its renormalized counterpart in the $\RI$ scheme  
\begin{equation}\label{eq:basic0}
  \mathcal{O}^{\RI}_{\mathrm{R}} (\mu) = Z^{\RI,
    \mathrm{lat}}_{\mathcal{O}}(\mu, a) 
  \mathcal{O}^\mathrm{lat}_\mathrm{bare}(a)\,. 
\end{equation}
The matching scale $\mu$ is restricted by two conditions: It should be
sufficiently large to avoid non-perturbative effects and small enough to avoid
large cut-off effects.  
For the scalar density, the non-perturbative matching was done in
\cite{Maillart:2008pv}, using a technique which removes a large part of
$\mathrm{O}(a)$ cut-off effects. 

In the second step one relates the renormalized operators in the $\RI$ and
$\MS$ schemes in perturbation theory  
\begin{equation}\label{eq:basic1}
  \mathcal{O}^{\MS}_{\mathrm{R}}(\mu) = Z^{\MS, \RI} (\mu)
  \mathcal{O}^{\RI}_{\mathrm{R}} (\mu)\,. 
\end{equation}
Obviously, the value of $\mu$ must lie in the perturbative regime. The factor
$Z^{\MS, \RI} (\mu)$ has been calculated up to NNLO \cite{Franco:1998bm} and
NNNLO \cite{Chetyrkin:1999pq}. 
Combining \eqref{eq:basic0} and \eqref{eq:basic1}, we obtain the matching
factor connecting the bare lattice and the renormalized $\MS$ results 
\begin{equation}\label{eq:basic2}
  \mathcal{Z}^{\MS, \mathrm{lat}}_{\mathcal{O}} (\mu, a) = Z^{\MS, \RI} (\mu)
  Z^{\RI, \mathrm{lat}}_{\mathcal{O}}(\mu, a) \,. 
\end{equation}

As the expansion parameter in the perturbative series is the running coupling
$\alpha(\mu)$ of QCD, we need to determine it for the scale $\mu$ where the
matching is done.   
The running of $\alpha$ is described by the differential equation 
\begin{equation}\label{eq:a}
  \frac {d \alpha(\mu)} {d \,\mathrm{ln} \, \mu^2} =  \beta(\alpha)  \,,
\end{equation}
where the 4-loop $\beta$-function in $\MS$ scheme is given in
\cite{vanRitbergen:1997va}. As initial condition we use $\alpha (2 \G) =
0.2904 $ from \cite{Aoki:2007xm} \footnote{This result is obtained by using
  the PDG value $\alpha (\mathrm{M_Z}) = 0.1176 \pm 0.002$  \cite{Yao:2006px}
  and running it from 5-flavors down to 3-flavors, across the $\mathrm{m_b}$
  and $\mathrm{m_c}$ thresholds}.The procedure described above is carried out
at several matching scales $\mu \geq 2 \G$.

In the third step Renormalization Group technique is used to obtain the
desired conversion factor 
$\mathcal{Z}^{\MS, \mathrm{lat}}_{\mathrm{s}}(\mu_0, a)$ 
for the scalar density at $\mu_0 = 2 \G$, 

\begin{equation}\label{eq:anomalous3}
 \mathcal{Z}^{\MS, \mathrm{lat}}_{\mathrm{s}} (\mu_0, a) = 
 \mathcal{Z}^{\MS,\mathrm{lat}}_{\mathrm{s}} (\mu, a) \, \mathrm{exp} 
 \left(- \int_{\alpha(\mu_0)}^{\alpha(\mu)} d\alpha \frac
   {\gamma_s(\alpha)}{\beta(\alpha)}\right)  \,. 
\end{equation}
Here, due to the relation $Z_s = Z_m^{-1}$, the anomalous dimension $\gamma_s$
is related to the mass anomalous dimension by $\gamma_\mathrm{s} = -
\gamma_\mathrm{m}$. The latter has been calculated to four loops in
\cite{Chetyrkin:1997dh, Vermaseren:1997fq}. 

\begin{figure} [h!t]
  \begin{center}
    \includegraphics[width=100mm]{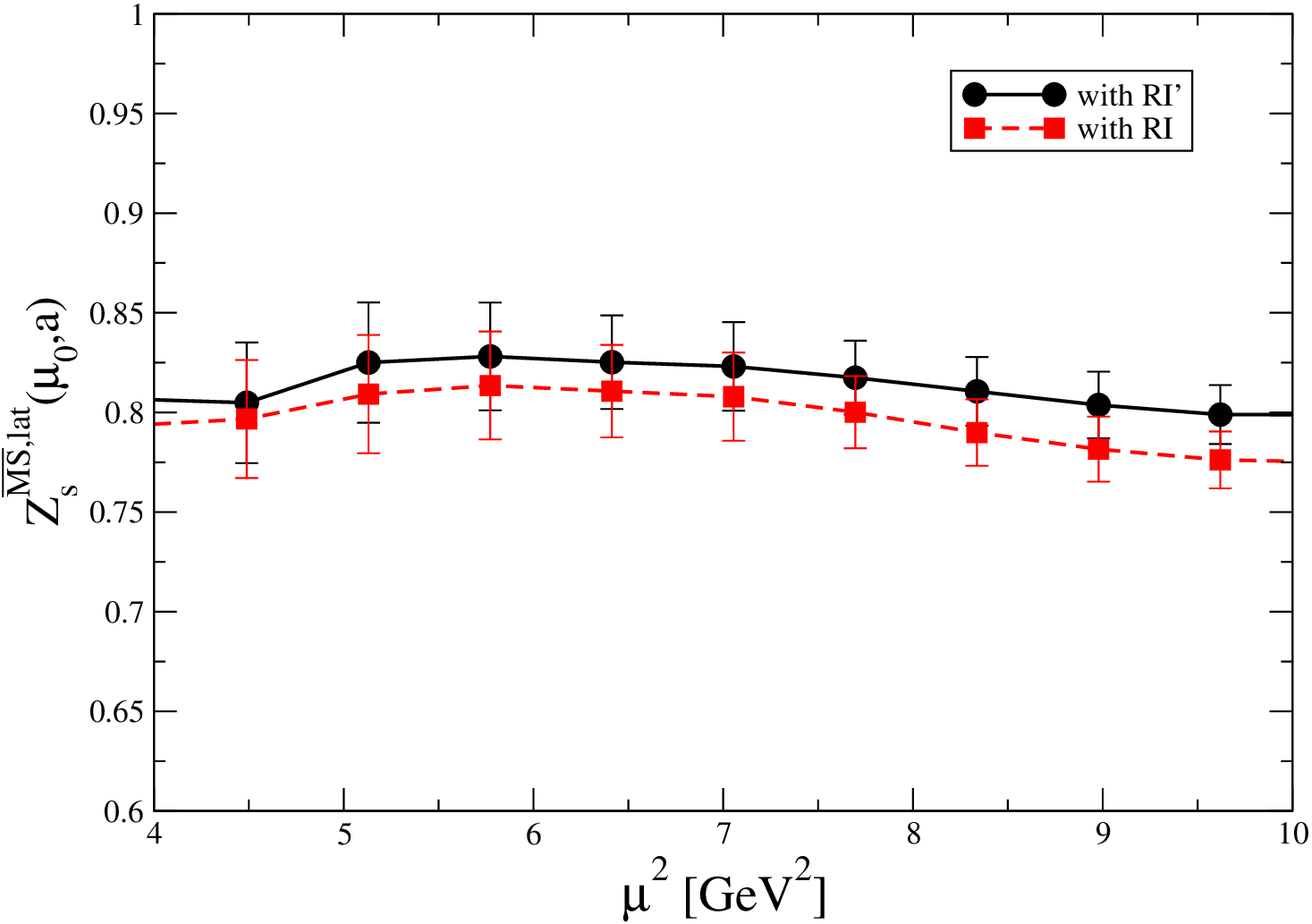}
    \caption{ {}The renormalization factor $\mathcal{Z}^{\MS,
        \mathrm{lat}}_{\mathrm{s}} (\mu_0, a)$ for the scalar density,
      connecting the $\MS$ and lattice schemes, at $\mu_0 = 2 \G$, obtained
      using $\RI$ and $\RI'$ schemes at the intermediate step vs. the matching
      scale $\mu$.} 
    \label{fig:Zs} 
  \end{center}
\end{figure}

In Fig.~\ref{fig:Zs}, we plot our result 
$\mathcal{Z}^{\MS, \mathrm{lat}}_{\mathrm{s}} (\mu_0, a)$ 
at $\mu_0 = 2 \G$ vs. the matching scale $\mu$ for the intermediate 
schemes $\RI$ and $\RI'$. 
If the non-perturbative, cut-off and higher order perturbative effects were
negligible, the two curves would coincide and would be independent of the
choice of $\mu$. 
Taking the average values for $ 2 \G \leq \mu \leq 3 \G$ and taking into
account the slight difference between the curves, we get  
\begin{equation}\label{eq:Z}
  \mathcal{Z}^{\MS, \mathrm{lat}}_{\mathrm{s}} (\mu_0 = 2 \G, a) = 0.82(3)
\end{equation}

\eject

\end{document}